# Prediction of fundamental properties of Be-B-Ta based novel ternary compounds from first-principles calculations

Enamul Haque[1*], M. Anwar Hossain[1], and Catherine Stampfl[2]

[1] Department of Physics, Mawlana Bhashani Science and Technology University, Santosh, Tangail-1902, Bangladesh

[2] School of Physics, The University of Sydney, Sydney, New South Wales, 2006, Australia.

*enamul.phy15@yahoo.com

**Abstract**

Be-B/B-Ta based compounds are very attractive to researchers because of their high density and ultra-hardness. But ternary Be-B-Ta compounds are neither synthesized nor predicted. In this paper, variable composition evolutionary crystal structure prediction calculations based on first-principles method have been performed to find the stable crystal structure containing Be-B-Ta at ambient condition. The predicted five compounds $BeB_2Ta$, $BeB_3Ta_2$ (high-pressure phase), $BeBTa$, $BeBTa_2$, and $Be_2B_2Ta$ have been found to be highly dense and very hard materials. All these compounds are metallic and spin-orbit coupling (SOC) effect is significant in them. Only two of them ($BeB_2Ta$ and $Be_2B_2Ta$) have been found to be superconductors within Migdal-Eliashberg theory. The calculated critical temperature including SOC effect is 8 and 9 K for $BeB_2Ta$ and $Be_2B_2Ta$, respectively. Because of their energetic and dynamic stability, these compounds might be favorable to synthesize in the laboratory.

## 1. Introduction

Beryllium (Be) and boron (B) are light elements with different electronic structures, first one is the metal and the second one is the semiconductor, while tantalum is the electron-rich transition metal. Beryllium has applications in X-ray equipment [1], as effective moderators and reflectors for neutrons [2], aircraft, rockets, etc. [3,4]. The use of a glove box is very effective way to synthesize industrially beryllium-containing compounds [5]. Different phases are possible by the combination of Be and B, such as $Be_2B$ [6], $Be_4B$ [7], and boron-rich compounds $BeB_2$ [8–12] and $BeB_4$ (high-pressure phase) [10]. Usually, covalent-bond forming element based compounds show high hardness, for example, $BeB_2$ (Cmcm) [9], diamond [13], c-BN [14], and $BC_2N$ [15]. The diamond, c-BN, and $BC_2N$ would be synthesized at high pressure and high-temperature condition [16]. On the other way, light element combined with electron-rich transition metal-based materials show also high hardness [17–21], such as $ReB_2$ [22], $IrN_2$ [23], $RuO2$ [24], $WB_4$ [25], $MnB_4$ and $CrB_4$ [26]. This method is very effective to design superhard materials. The different phases of tantalum borides (TaB, $TaB_2$, $Ta_3B_4$) have been synthesized at ambient pressure [27]. It has been found that TaB and $TaB_2$ have high hardness (~30 GPa) [27–29]. The theoretically calculated Vickers hardness of TaB is 28.6 GPa [30]. Thus, transition metal rich borides show good mechanical behavior for practical applications.

Boron-rich $BeB_2$ and $MgB_2$ shows superconductivity, although the first one has very low transition temperature (0.72 K) [8], the second one has the highest transition temperature (39 K) among simple binary structure [31]. In $MgB_2$, σ-band electrons strongly bonded with in-plane phonons to induce a high superconducting transition temperature. Boron-rich transition metal boride $TaB_2$ might show superconductivity with a critical temperature of 9.5 K [32,33], although others reported the absence of superconductivity [34,35]. The strong hybridization between Ta 5d and B 2p orbitals outside the hexagonal basis plane might be responsible for the absence of superconductivity. So, Be 2s may change the nature of hybridization and brings the superconductivity. However, no one has been reported it yet. The ternary compounds containing Be-B-Ta are still uninvestigated. From the above points of view, it is very interesting to study the possible crystal structure containing Be-B-Ta and their fundamental properties, such as electronic structure and superconducting properties.

In this work, variable-composition evolutionary crystal structure search calculations have been performed to find thermodynamically stable structure containing Be-B-Ta. The predicted $BeB_2Ta$ (monoclinic), $BeB_3Ta_2$ (body-centered orthorhombic), BeBTa (hexagonal), $BeBTa_2$ (base-centered orthorhombic), and $Be_2B_2Ta$ (body-centered tetragonal), have been found to be highly dense and very hard materials. These compounds show metallic band structure and spin-orbit coupling (SOC) effect is significant in them. Only boron-rich ($BeB_2Ta$ and $Be_2B_2Ta$) have been found to be superconductors within Migdal-Eliashberg theory. Tantalum rich or equivalent stoichiometry compounds do not show superconductivity because of the dominant contribution of Ta-5d.

## 2. Results and Discussion
## 2.1. Structural properties

The ternary crystal structure search reveals many possible phases of Be-B, Be-Ta, B-Ta, and Be-B-Ta, as plotted in ternary convex hull diagram in Fig. 1. The corresponding formation enthalpies are shown in the left panel.

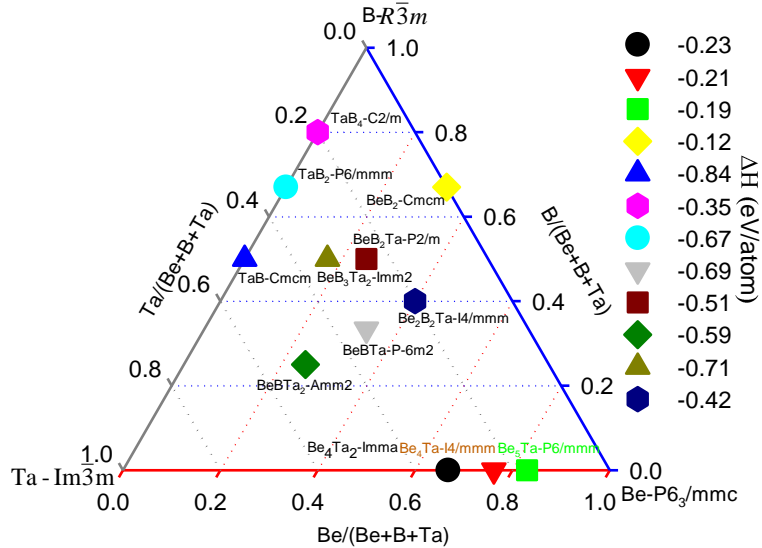

Fig. 1: Ternary convex hull (composition- formation enthalpy) diagram of all stable compounds containing Be-B-Ta at ambient condition. The right legend represents the value of formation enthalpy (eV/ atom) with the corresponding symbol in the convex hull diagram. The stable phases with lower negative formation enthalpy are indicated by the corresponding symbols closer to the convex hull.

The calculated phase of hexagonal beryllium (P63/mmc, #194) [36,37], α-rhombohedral boron ($R\bar{3}m$, #160) [38], and body-centered cubic tantalum ($Im\bar{3}m$, #229) [39] are consistent with the available data. The boron rich $BeB_2$ (Cmcm, #63) is a thermodynamically stable structure, as reported in the Ref. [9,12]. The present calculations also reproduce the previously reported stable structure TaB (Cmcm,

#63) [27], TaB$_2$ (P6/mmm,#191)[27], and TaB$_4$ (C2/m, #12) [40]. Besides these phases, beryllium rich Be$_4$Ta$_2$ (Imma, #74), Be$_4$Ta (I4/mmm, #139), and Be$_5$Ta (P6/mmm, #191) are uncovered for first time. These phase are thermodynamically stable at ambient condition. However, several intermediate phases of Be-Ta) during solidification reported in Ref. [41] are not found in the present calculations. These beryllium rich phases will not be focused in this paper. The present crystal structure searches uncover five ternary stable phases containing Be-B-Ta. Among them, BeB$_2$Ta and BeB$_3$Ta$_2$ crystallize in a monoclinic with unique-axis c and body-centered orthorhombic structure, respectively, while tantalum rich BeBTa$_2$ forms base-centered orthorhombic structure.

Table-1: The fully relaxed lattice parameters, space group symbol (number), and atomic density of the stable ternary compounds containing Be-B-Ta. The fully relaxed atomic coordinates are provided in the the supporting information (SI).

| Compound | Space group | a (Å) | b (Å) | c (Å) | ρ (g/cc) |
| --- | --- | --- | --- | --- | --- |
| BeB$_2$Ta | P2/m(c) (#10) | 7.2382 | 3.1629 | 3.2027 (γ=102.55) | 9.77 |
| BeB$_3$Ta$_2$ | Imm2 (#44) | 6.5468 | 3.1666 | 5.5247 | 12.98 |
| BeBTa$_2$ | Amm2(#38) | 3.2582 | 3.3278 | 8.9107 | 13.18 |
| BeBTa | P$\bar{6}$m2 (#187) | 3.2629 | 3.2629 | 3.2404 | 11.18 |
| Be$_2$B$_2$Ta | I4/mmm (#139) | 3.1808 | 3.1808 | 8.5602 | 9.56 |

With 1:1:1 stoichiometry, BeBTa forms a primitive hexagonal structure. Both beryllium and boron-rich Be$_2$B$_2$Ta crystalizes in body-centered tetragonal structure. The fully relaxed lattice parameters and density of these phases are listed in Table 1. The calculated density of these phases suggests high hardness of them. The ground state crystal structure of these compounds are shown in Fig. 2. BeB$_2$Ta forms AgPtO$_2$-type structure with one B-Ta layer [42].

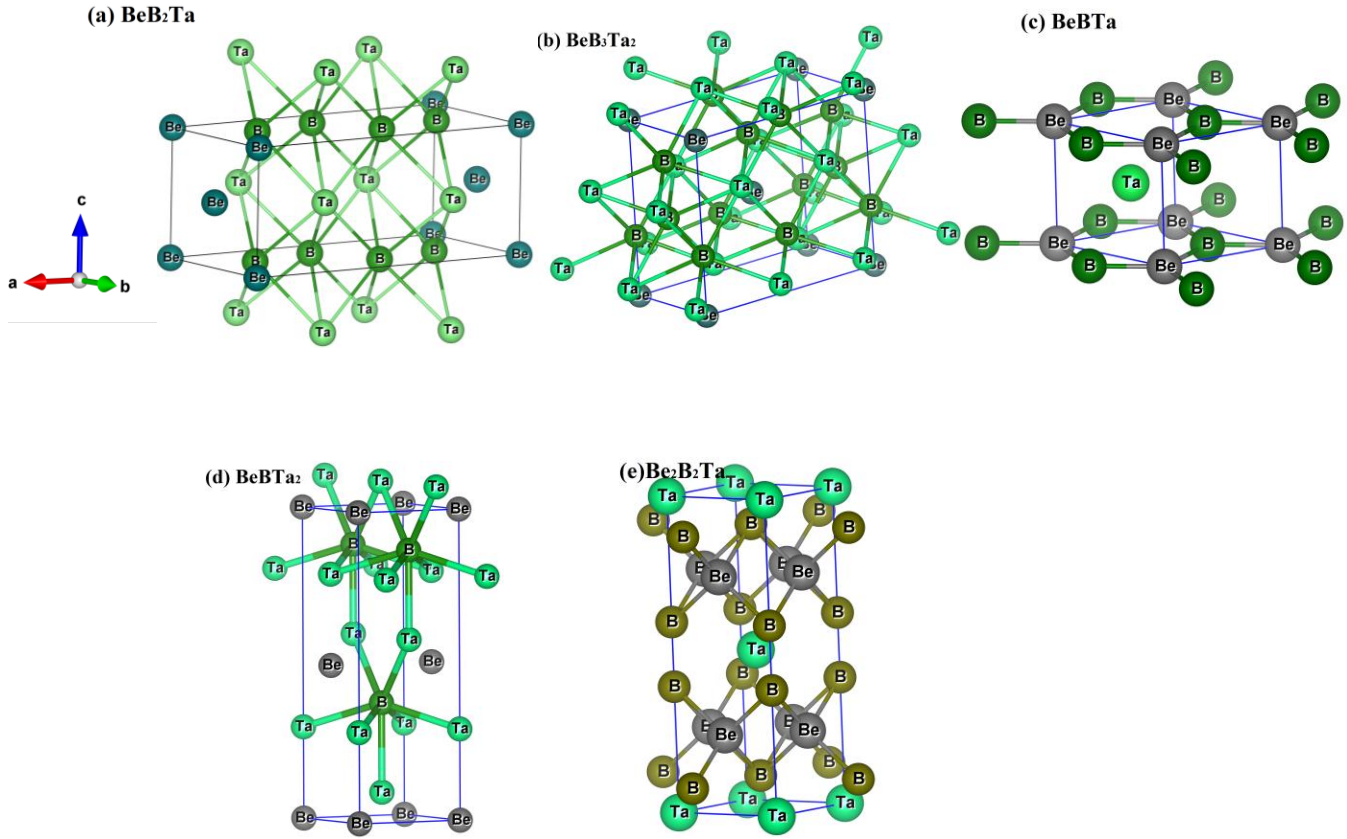

Fig. 2: The ground state crystal structure of: (a) Monoclinic (with unique axis-c) $BeB_2Ta$, (b) body-centered orthorhombic $BeB_3Ta_2$, (c) hexagonal BeBTa, (d) base-centered orthorhombic $BeBTa_2$, and (e) body-centered tetragonal $Be_2B_2Ta$.

The orthorhombic $BeB_3Ta_2$ forms $NaPd_3Si_2$ type structure [43] while BeBTa forms a hexagonal YbPtP type structure [44]. The tantalum-rich $BeBTa_2$ is a layered orthorhombic structure, like $CeNiC_2$ [45]. Both beryllium and boron-rich $Be_2B_2Ta$ is a layered tetragonal structure. This is similar to the structure of $ThCr_2Si_2$ [46]. The structure of $Be_2B_2Ta$ consists of a stack of $Be_2B_2$ layers. Four Be atoms surround four B atoms. On the other hand, B atom has four Be nearest neighbors and one B atom closet to it. The combination of them forms a tetragonal pyramid, as shown in Fig. 2(e). In $BeB_2Ta$, $BeB_3Ta_2$, and $BeBTa_2$, B and Ta atoms are bonded together while Be and B are bonded together in BeBTa and $Be_2B_2Ta$. To take into account the mechanical stability, the calculated elastic constants of these compounds are listed in Table 2. The listed elastic constants fulfill the mechanical stability criteria for these systems described in Ref. [47]. Thus, they are mechanically stable crystal structures. The real phonon frequencies ensure the dynamical stability a crystal system. The calculated phonon dispersion relations of these compounds with and without spin-orbit coupling effect have no negative frequencies, as shown in the left panel of Fig. 5-6. Therefore, all these phases are dynamically stable at ambient condition. These results suggest that they are favorable to form at ambient condition. The critical value of Pugh's ratio of 1.75 differentiate the brittle and ductile materials [48]. The low value of Pugh's ratio for these compounds indicates their brittle nature.

Table-2: Calculated elastic constants, bulk and shear modulus (in GPa), Pugh's ratio, and theoretical Vicker's hardness (see the computational methodology) (GPa) of all stable ternary compounds.

| Parameter | $BeB_2Ta$ | $BeB_3Ta_2$ | $BeBTa_2$ | $BeBTa$ | $Be_2B_2Ta$ |
|---|---|---|---|---|---|
| $c_{11}$ | 489.40 | 493.03 | 448.63 | 491.89 | 529.43 |
| $c_{12}$ | 107.32 | 146.20 | 187.01 | 128.72 | 116.45 |
| $c_{13}$ | 96.97 | 131.40 | 127.93 | 123.49 | 103.24 |
| $c_{15}$ | -1.81 | --- | --- | --- | --- |
| $c_{22}$ | 554.82 | 590.28 | 441.35 | --- | --- |
| $c_{23}$ | 155.52 | 95.73 | 115.31 | --- | --- |
| $c_{25}$ | -0.038 | --- | --- | --- | --- |
| $c_{33}$ | 536.08 | 596.91 | 508.54 | 513.50 | 543.54 |
| $c_{35}$ | -0.042 | --- | --- | --- | --- |
| $c_{44}$ | 236.31 | 244.56 | 199.22 | --- | 193.92 |
| $C_{46}$ | -1.78 | --- | --- | --- | --- |
| $c_{55}$ | 154.54 | 238.65 | 179.01 | 251.07 | --- |
| $c_{66}$ | 148.02 | 244.30 | 244.18 | ---- | 249.46 |
| B | 253.34 | 269.32 | 250.96 | 249.83 | 249.80 |
| G | 185.98 | 231.18 | 185.03 | 209.14 | 211.82 |
| B/G | 1.37 | 1.16 | 1.35 | 1.19 | 1.18 |
| $H_v$ | 26.62 | 37.39 | 26.68 | 34.0 | 34.84 |

Since these materials are highly dense, it is reasonable to calculate Vickers hardness by using Chen's formula [49]. This formula gives an idea about the hardness of a material. The calculated values of Vickers hardness are presented in Table 2. All the phases have very high hardness and $BeB_3Ta_2$ has the highest hardness among them. The hardness of $BeB_3Ta_2$ is close to the critical value (40 GPa) of a superhard material. Therefore, the calculated hardness suggests that these compounds are potential candidates for superhard materials.

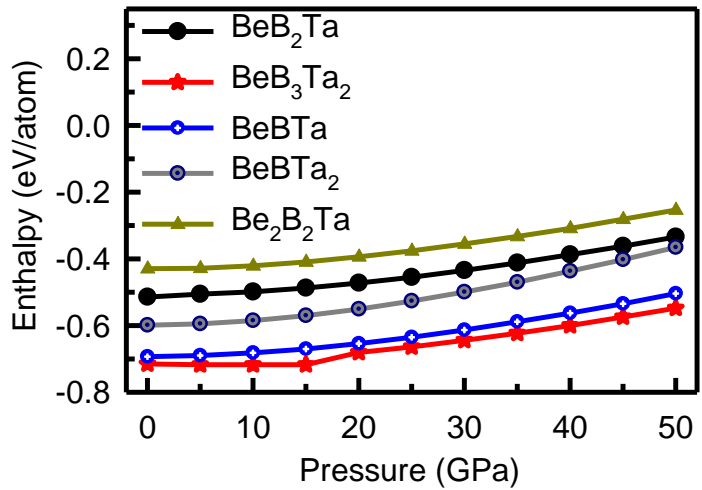

Fig. 3: Pressure-dependent formation enthalpy of the predicted Be-B-Ta containing phases. The negative value of the formation enthalpy indicates that the corresponding phase is thermodynamically stable and probable to form.

The calculated pressure dependent formation enthalpy of all stable phases is shown in Fig. 3. All the predicted phases are thermodynamically stable up to the studied pressure range. It is interesting to note that $BeB_3Ta_2$ is more favorable to form at 15 GPa than that at the ambient condition. Therefore, it can be regarded as a high-pressure phase.

## 2.2 Electronic properties

Since the predicted five compounds are stable, it is safe to proceed the description their electronic structure. The calculated electronic band structure, total and projected density of states of $BeB_2Ta$ and $BeB_3Ta_2$ are shown in Fig. 4. The gross features of the band structures of these two compounds are completely different, although they contain the same elements. One valence band and one conduction bands of BeB$_2$Ta cross the Fermi level. The valence band arises from B 2p states, and one conduction band arises from Ta 5d states. All these bands are highly dispersive. A pseudogap appears at X-point from B 2p states and the Fermi level lies at the middle of the pseudogap. Interestingly, another pseudogap opens up at C-point from dominant Ta 5d and minor Be 2p states. In that case, the Fermi level lies at the top of the pseudogap. Although, Ta 5d states have the dominant contribution to the density of states at the Fermi level, both Be 2p and B 2p states (especially B 2p) have also significant contributions. Thus, Ta 5d orbitals are strongly hybridized with Be 2p and B 2p, unlike with B 2p orbitals only in TaB2 [34]. This indicates the possibility of superconductivity in BeB$_2$Ta. Only few bands arisen from Ta 5d states are split due to the inclusion of spin-orbit coupling effect, as presented by dark yellow dash lines in Fig. 4. The density of states including spin-orbit coupling effect is not presented here for clarity. Unlike BeB$_2$Ta, six conduction bands of BeB$_3$Ta$_2$ cross the Fermi level, where most of the bands arise from Ta 5d states. A pseudogap opens along T-R directions below the Fermi level from B 2p states.

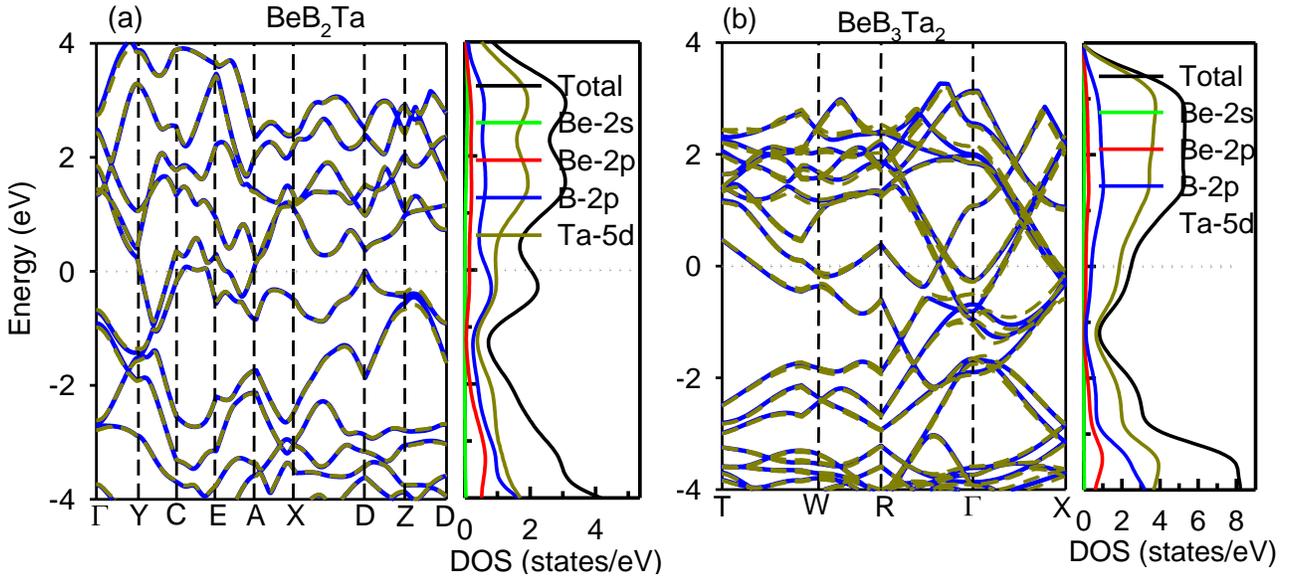

Fig. 4: Electronic band structure, total and projected density of states (right panel) of: (a) BeB$_2$Ta and (b) BeB$_3$Ta$_2$. The dark yellow dash lines represent energy bands calculated including spin-orbit coupling effect. The dotted line at zero energy represents the Fermi level. The density of states with SOC effect is not shown here for clarity of the figure.

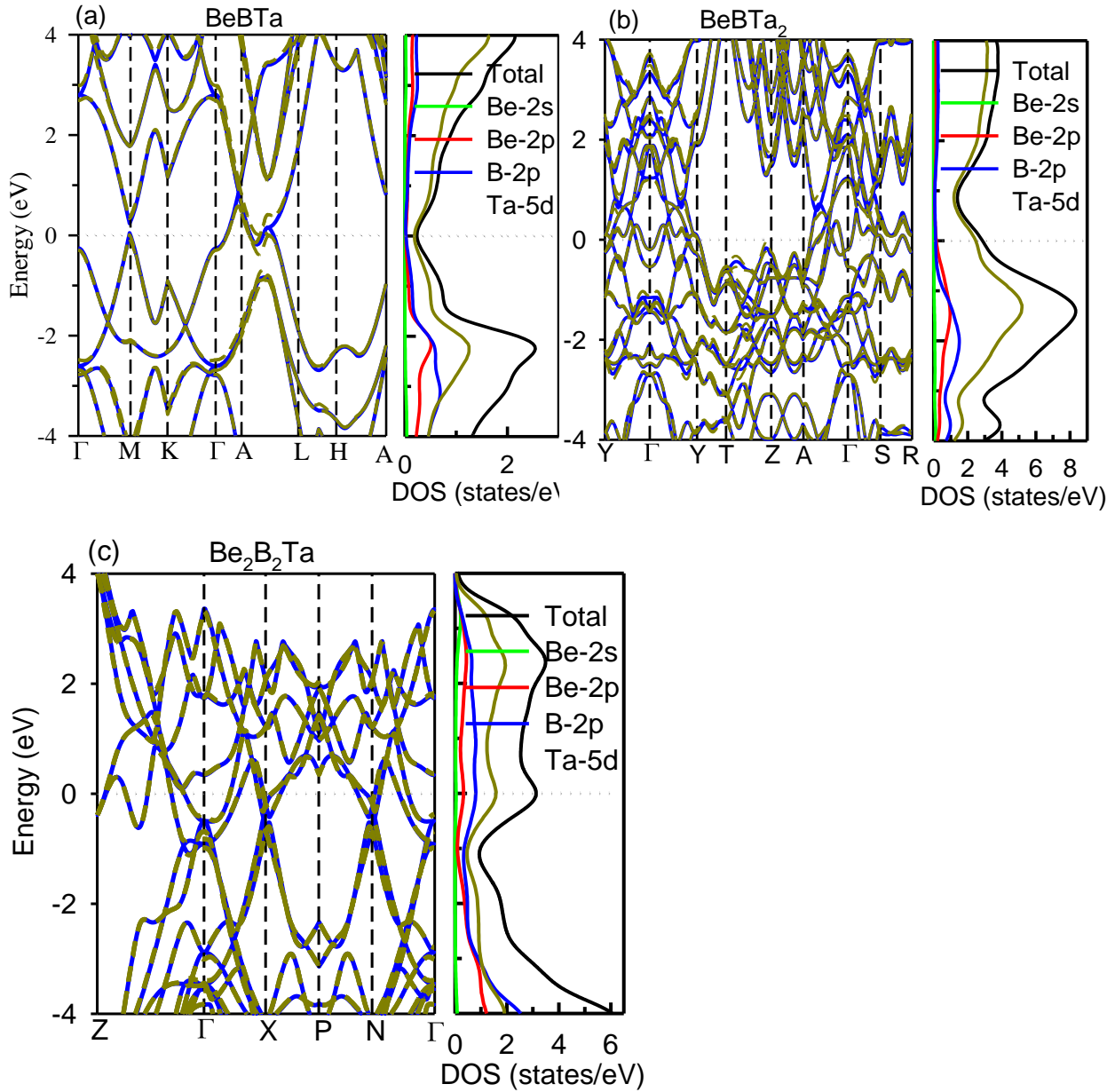

Fig. 5: Electronic band structure, total and projected density of states (right panel) of: (a) BeBTa, (b) BeBTa$_2$, and (c) Be$_2$B$_2$Ta. The dark yellow dash lines represent energy bands calculated including spin-orbit coupling effect. The dotted line at zero energy represents the Fermi level. The density of states with SOC effect is not shown here for clarity of the figure.

From the calculated projected density of states, it is clear that Be 2p and B 2p states have not significant contribution to the density of states at the Fermi level. Furthermore, Ta 5d states are strongly hybridized with B 2p. Therefore, BeB$_3$Ta$_2$ may not be a superconducting material. Like BeB$_2$Ta, certain bands arisen from Ta 5d states are split by spin-orbit coupling effect, as indicated by dark yellow dash lines. The valence bands and conduction bands overlap in both compounds, hence, they are conductors at ambient condition.

Unlike above two compounds, only two bands of hexagonal BeBTa, arisen from Ta 5d, cross the Fermi level and another band only touch the Fermi level at M-point. Neither Be 2p nor B 2p states have any contribution in the electronic structure around the Fermi level. Three pseudogaps along Γ-A and L-A directions open from B 2p states. The Fermi level lies at the middle of these pseudogaps. It is clear that the calculated total density of states at the Fermi level is very small and BeBTa can be regarded as a carrier deficient compound. Another fact that PBE functional underestimates the electronic bandgap 30-100% by experimental value [50,51], this compound may be a semimetal. Note that the band structure of BeBTa has some features of Weyl semimetal [52]. Further studies are required to clarify these results. The band structure of tantalum-rich BeBTa$_2$ is totally different from that of the first three phases. Total eight bands cross the Fermi level and all these bands arise from Ta 5d states. A large pseudogap opens along Y to A direction. The Fermi level lies at the bottom of the pseudogap. This pseudogap arises from overlapping between Be 2p and B 2p states. The density of states at the Fermi level of Be 2p and B 2p orbitals is negligibly small as compared to that of Ta 5d states. Therefore, BeBTa$_2$ may not be a superconductor. Note that both conduction and valence bands are split by the spin-orbit coupling effect. Although Be 2p and B 2p electrons have contributions in valence bands, they have not significant contributions in conduction bands, as can be clearly seen from the calculated projected density of states. In contrast to BeBTa$_2$, both Be and B rich Be$_2$B$_2$Ta is a metal with two bands crossing the Fermi level. One band originates from Ta 5d states and another band from the overlapped between Be 2p and B 2p states. Both Be 2p and B 2p states have significant contributions to the Fermi level. This can be clearly seen from the projected density of states as shown in the left panel of Fig. 5 (c). A pseudogap opens along X-N direction and the Fermi level lies at the top of the pseudogap. This pseudogap arises from Ta 5d and Be 2p states. From the calculated density of states, it is clear that total density of states at the Fermi level is much higher than that of BeB$_2$Ta. The spin-orbit coupling is less significant in Be2B2Ta because Ta is not dominant in this phase. Only few bands are split by the spin-orbit coupling effect. Since Ta 5d orbitals are hybridized with Be 2p and B 2p states, this phase may be a potential two bands superconductor. Furthermore, all these bands are highly dispersive. The conduction bands and valence bands overlap each others indicating its metallic nature. Therefore, these materials would be very useful in different technological applications, for example as a superhard metallic conductor. Further theoretical studies and experimental synthesis of these phases are encouraged to clarify these predictions.

## 2.3 Superconductivity

The calculated electronic structure predicts that two phases of Be-B-Ta may be potential superconductors. Moreover, $MgB_2$, a simple binary, shows high-temperature superconductivity [31], although there are controversies about the superconductivity of $TaB_2$ [32,34]. Therefore, it is interesting to study the electron-phonon interactions in Be-B-Ta containing phases. The calculated phonon dispersion of $BeB_2Ta$ and $BeB_3Ta_2$ is shown in the left panel of Fig. 6. The lower frequency phonons (acoustic phonons) of $BeB_2Ta$ arise from Ta atoms while phonons within the frequency range 300-600 $cm^{-1}$ arise from Be and B. Both Be and B give rise to the higher frequency optical phonons. It is interesting to note that constituent elements separate the phonon band structure into three regions. The phonon bands of one region do not overlap with the phonon bands of other regions. Two bands within the frequency range ~300-600 $cm^{-1}$ correspond to vibrations of B and Be in opposite directions. These three regions induce three dominant peaks in the phonon density of states, as shown in the middle panel of Fig. 6(a). Therefore, three elements contribute to the phonons almost equally. Unlike $BeB_2Ta$, the phonon band structure of $BeB_3Ta_2$ has four regions (Left panel of Fig. 5(b)). The acoustic phonons arise from Ta atoms, optical phonons within the frequency range 400-600 $cm^{-1}$ from B atoms, and higher frequency optical phonons (600-900 cm-1) arise from both Be and B. Like $BeB_2Ta$, no phonon bands of one region overlap with the phonon bands of other regions. But the optical phonons of two regions around 550 $cm^{-1}$ overlap with each other when the spin-orbit coupling effect is included (dark yellow line in the dispersion curve). These four regions corresponding to the four dominant peaks in the atom projected phonon density of states, as shown in the middle panel of Fig. 6 (b). A large gap exists within the frequency range 200-400 $cm^{-1}$. It is clear that the phonon density of states of $BeB_3Ta_2$ is much higher than that of BeB2Ta. Some of the phonon bands of both compounds are split by the spin-orbit effect. However, this effect is much sensitive in Ta-rich $BeB_3Ta_2$ than that in $BeB_2Ta$. Such an effect has been found in many elemental superconductors [53]. No imaginary frequencies in the phonon dispersion ensure the dynamical stability of both compounds.

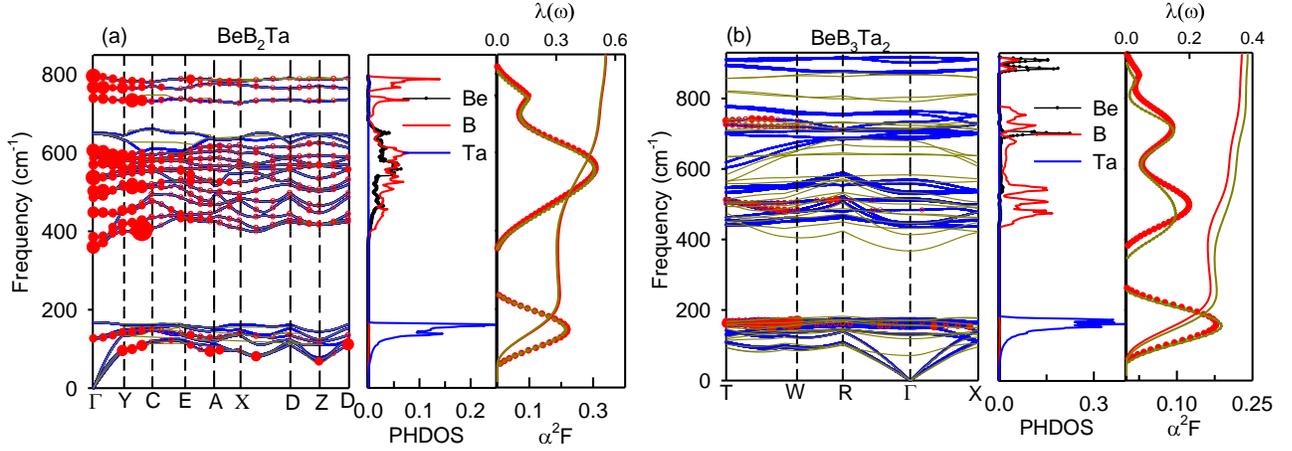

Fig. 6: Electron-phonon related properties of: (a) BeB$_2$Ta and (b) BeB$_3$Ta$_2$. The left panel shows phonon dispersion relations and middle panel shows atom projected phonon density of states. The dark yellow line in the phonon band structure represents the spin-orbit coupling effect included data. The phonon density of states including spin-orbit coupling effect is not shown here for clarity. The right panel shows Eliashberg spectral functions and integrated electron-phonon coupling constant. Similarly, the red circle-symbolized line represents Eliashberg spectral function without spin-orbit coupling effect while dark yellow star-symbolized line represents the Eliashberg spectral function including SOC effect. The red circles represent mode dependent electron-phonon coupling constant over the Brillouin zone (see the text).

In the Migdal-Eliashberg formulism [54], the Eliashberg spectral function is written as [55,56]

$$\alpha^2 F(\omega) = \frac{1}{2\pi N(E_F)} \sum_{qv} \delta(\omega - \omega_{qv}) \frac{\gamma_{qv}}{\hbar \omega_{qv}} \dots \dots \dots \dots (1)$$

where $N(E_F)$ is the density of states at the Fermi level and $\gamma_{qv}$ is the electron-phonon linewidth for wave vectors $q$ and $v$.

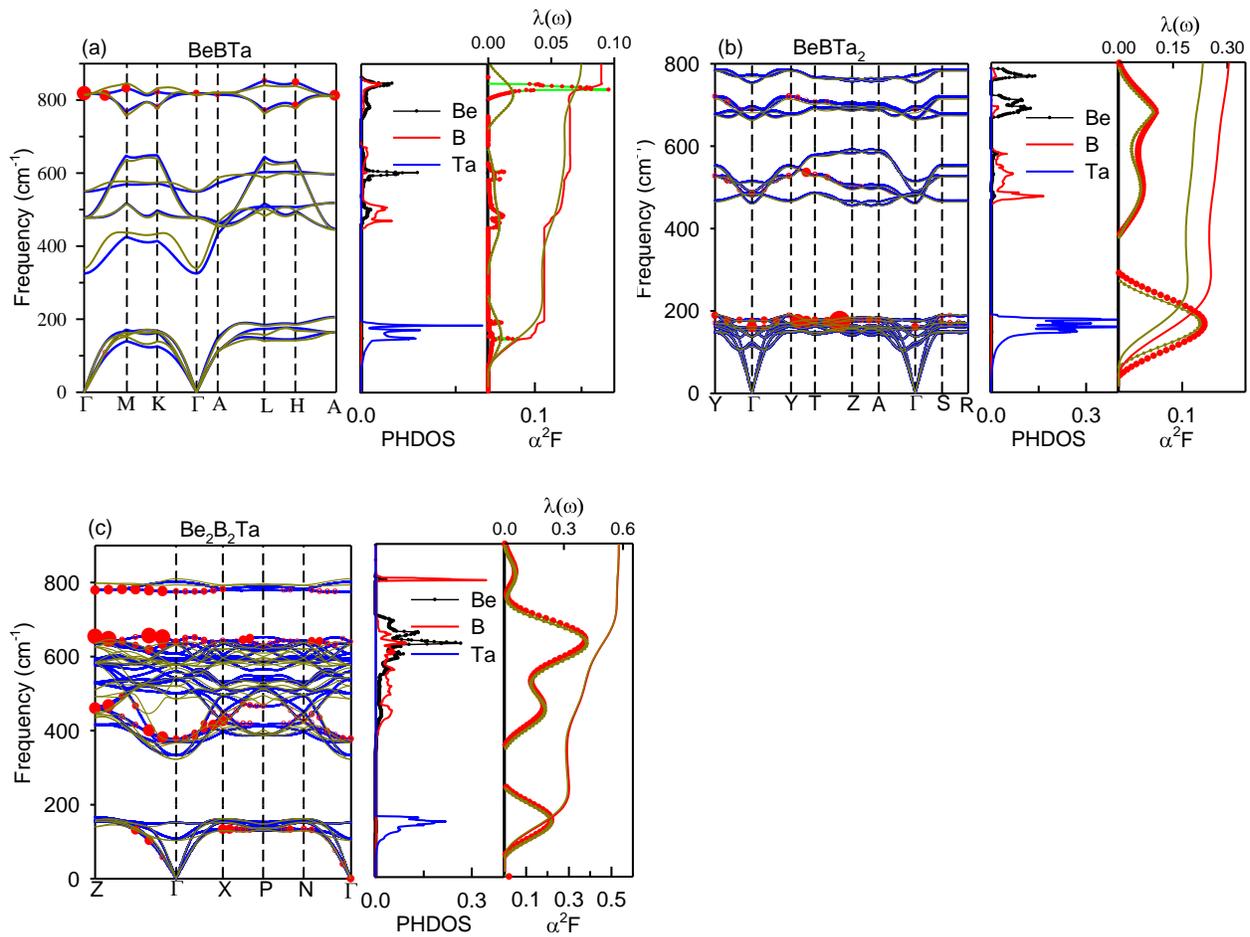

Fig. 7: Electron-phonon related properties of: (a) BeBTa, (b) $BeBTa_2$, and (c) $Be_2B_2Ta$. The left panel shows the phonon band structure and middle panel shows atom projected phonon density of states. The dark yellow line in the phonon band structure represents the spin-orbit coupling effect included data. The phonon density of states including spin-orbit coupling effect is not shown here for clarity. The right panel shows Eliashberg spectral functions and integrated electron-phonon coupling constant. Similarly, the red circle-symbolized line represents Eliashberg spectral function without spin-orbit coupling effect while dark yellow star-symbolized line represents the Eliashberg spectral function including SOC effect. The red circles represent mode dependent electron-phonon coupling constant over the Brillouin zone (see the text).

The calculated Eliashberg spectral function of both compounds is shown in the right panel of Fig. 6 with and without spin-orbit coupling effect (dark yellow star-symbolized line and red circle-symbolized line, respectively). The Eliashberg function of both compounds has three dominant peaks, although phonon density of $BeB_3Ta_2$ has four dominant peaks. The highest peak in Eliashberg function (EF) of $BeB_2Ta$ arises from B atom while it arises from Ta in the case of $BeB_3Ta_2$. From the comarison of EF with the phonon density of states of $BeB_2Ta$, it is clear that the electron-phonon (e-ph) interactions are almost equally distributed among all of the phonon modes. But e-ph interactions in $BeB_3Ta_2$ are distributed only among some phonon branches. The higher frequency optical phonons of $BeB_2Ta$ have significant contribution to the Eliashberg spectral function, but they have negligble contribution to the EF in the case of $BeB_3Ta_2$. Low frequency acoustic phonons (0-200 cm$^{-1}$) give rise to ~50 % of the total electron-phonon coupling (EPC) in $BeB_3Ta_2$ (see the right panel of Fig. 5 (b)). But only ~25% of the total EPC comes from acoustic phonons in $BeB_2Ta$ while optical phonons (from Be and B) provide 75 % of the total EPC. This can be further understood from the mode dependent EPC calculations. The mode dependent electron-phonon coupling constant, i.e., the partial contribution of the each phonon mode to total electron-phonon coupling constant can be written as

$$\lambda_{qv} \equiv \frac{1}{\pi N(E_F)} \frac{\gamma_{qv}}{\omega_{qv}^2} \dots\dots\dots\dots (2)$$

By taking summation of $\lambda_{qv}$ over $v$ and average on the Brillouin zone, total electron-phonon coupling constant (EPC) λ can be be calculated. The EPC is determined by [54,56]

$$\lambda = 2\int \frac{\alpha^2 F(\omega)}{\omega} d\omega \dots\dots\dots\dots (3)$$

The e-ph coupling distribution is shown in the phonon band structure (left panel of Fig. 6) by the red circles. The radius of the circle is directly proportional to the partial contribution of each phonon to the total e-ph coupling constant (λ). From the figure, the red circles are almost equally distributed among all of the phonon branches of $BeB_2Ta$ but some phonon branches of $BeB_3Ta_2$. The e-ph interactions around the intraband nesting point Γ are present. The strongest e-ph interactions of $BeB_2Ta$ are concentrated at Γ-C-points but they are concenered around T-W-points in $BeB_3Ta_2$. It is clear that Ta atoms (acoustic phonons) have the dominant contribution to the total e-ph constant of $BeB_3Ta_2$ but both Be and B have significant contribution to e-ph constant of BeB$_2$Ta. In a good electron-phonon superconductor, the e-ph coupling is usually concentrated in some phonon modes. Although BeB$_3$Ta$_2$ shows this type of distribution which can be seen from the almost perfect proportionality between phonon density of states and Eliashberg spectral function, the strength of e-ph coupling is very weak (small circle). But BeB$_2$Ta shows deviation, although the strength of the e-ph coupling is strong (large circle).

Unlike BeB$_2$Ta and BeB$_3$Ta$_2$, the hexagonal BeBTa and Ta-rich BeBTa$_2$ show little e-ph coupling. The phonon dispersion of BeBTa is similar to BeB$_2$Ta, it can be divided into three regions without any overlapping among phonon bands of each region. But acoustic and low-frequency optical phonons have almost negligible contribution in the e-ph coupling. Only few higher frequency optical phonons around 800 cm$^{-1}$ are strongly coupled with electrons. These phonons arise from Be and B where Be and B vibrate in opposite direction. The quite low density of states of B at the Fermi level, i.e., few available electrons of B, can only participate in the e-ph coupling. The corresponding Eliashberg function is quite low. EF contains only a dominant peak around 800 cm$^{-1}$. The SOC effect changes the Eliashberg spectral function and shifts the peaks to higher energy. Like phonon density of states, the gross features of EF are similar to the phonon density of states. The gross features of phonon dispersions of BeBTa$_2$ are similar to that of BeB$_3$Ta$_2$. A large gap exists between acoustic and optical phonons (from 200-500 cm$^{-1}$), as shown in the left panel of Fig. 7(b). The acoustic phonons arisen from Ta atom (see the atom projected phonon density of states) provide ~75% of total e-ph coupling. But the coupling strength is weak, like BeB$_3$Ta$_2$. The optical phonons arisen from B atoms have a little e-ph coupling, as it can be seen from the red circles in the phonon dispersion curve. Because Ta 5d and B 2p orbitals are strongly hybridized like that in TaB$_2$. Unlike the above two phases, the e-ph coupling of Be$_2$B$_2$Ta is almost equally distributed among all phonon modes. Like BeB$_2$Ta, the phonon band structure can be divided into three regions. A large gap exists between ~180-350 cm$^{-1}$. Acoustic phonons arisen from Ta have little contribution to the total e-ph coupling. The optical phonons within the frequency range ~350-620 cm$^{-1}$ have dominant contribution to the total e-ph coupling. These optical phonons arise from both Be and B atoms, where they move back and forth. Another peak in the atom projected phonon density of states of Be$_2$B$_2$Ta corresponds to the optical phonons arisen from B. Unlike BeB$_2$Ta, the strongest e-ph couplings (see red circles in the left panel of Fig. 7(c)) are concentred within Z to Γ point. The three dominant peaks in the phonon dos correspond to the three peaks in the Eliashbergh spectral function as shown in the right panel of Fig. 7(c). EF shows almost perfect proportionality with phonon dos, which is rare in some good e-ph superconductors. It is clear that SOC effect is less sensitive in this phase as compared to above-described phases. Only few phonon bands are split by SOC effect. The Eliashberg function of $Be_2B_2Ta$ remains almost same when SOC effect is included in the calculation, as Be and B have dominant contribution to e-ph coupling.

By using the calculated Eliashberg spectral function, the logarithmic average phonon frequency ($\omega_{ln}$) can be calculated by peforming the following numerical integration [54,56]

$$\omega_{ln} = \exp\left[\frac{2}{\lambda}\int \frac{d\omega}{\omega} \alpha^2 F(\omega) \ln(\omega)\right] \ldots \ldots \ldots \ldots (4)$$

The calculated total electron-phonon coupling constant and logarithmic average phonon frequency of these phases are listed in Table-3. The EPC of $BeB_3Ta_2$ and $BeBTa_2$ is very small while BeBTa has a negligible EPC. According to the electronic structure and e-ph properties analysis, these results are not surprising. Only two B-rich phases ($BeB_2Ta$ and $Be_2B_2Ta$) have a higher EPC as compared to others, as expected. In all phases except $BeB_2Ta$ and $BeB_3Ta_2$, the SOC effect reduces the EPC constant slightly. The logarithmic average phonon frequency of all compounds except $BeB_2Ta$ and $BeB_3Ta_2$ is increased when the SOC effect is included.

Using the calculated EPC and logarithmic average phonon frequency, the superconducting transition temperature can be calculated by using the Allen-Dynes equation [55,56]

$$T_c = \frac{\omega_{ln}}{1.2} \exp\left[\frac{-1.04(1+\lambda)}{\lambda(1-0.62\mu^*)-\mu^*}\right] \ldots \ldots \ldots \ldots (5)$$

where µ* stands for Coulomb pseudopotential constant and its value ranges between 0.1 and 0.15 [57,58]. Here, the most widely accepted value 0.1 of µ* has been used to evaluate superconducting transition temperature.

Table-3: The calculated total EPC ($\lambda$), logarithmic average phonon frequency ($\omega_{ln}$), and superconducting transition temperature of all stable phases at ambient condition. Inside the bracket, the calculated value of the parameter including SOC effect is presented. In all cases, the value 0.1 of $\mu^*$ is used.

| Compound | $\lambda$ | $\omega_{ln}$ (K) | Tc (K) |
| --- | --- | --- | --- |
| BeB$_2$Ta | 0.519 (0.526) | 553 (552) | 7.75 (8.13) |
| BeB$_3$Ta$_2$ | 0.338 (0.366) | 556.1(394.3) | 0.75(0.95) |
| BeBTa$_2$ | 0.265 (0.211) | 394.0 (565.42) | 0.05 (~0) |
| BeBTa | 0.081(0.076) | 411.9 (430.51) | ~0(~0) |
| Be$_2$B$_2$Ta | 0.546 (0.545) | 528.4 (539.74) | 8.9 (9) |

From the Table-3, it is clear that the superconductivity is absence in the Ta-rich (or equivalent) phases, BeBTa$_2$ and BeBTa. The BeB$_3$Ta$_2$ may not be a superconductor since the calculated Tc is too small. These results are consistent with the electronic structure calculations. Only BeB$_2$Ta and Be$_2$B$_2$Ta are superconductors. The obtained Tc of these two phases is slightly increased due to the inclusion of SOC effect, as the total EPC and logarithmic average phonon frequency are changed. Therefore, Be-B brings the superconductivity in these two phases of Be-B-Ta (see the supporting information (SI) for further discussion). The above results predict that any combination of B-rich with electron-rich transition metal may be potential superconductor if the hybridization between B-2p and electron-rich element-d orbitals is disturbed by another 2s-element.

## 3. Conclusions

In summary, the stable crystal structure containing Be-B-Ta at ambient condition has been predicted by using variable composition evolutionary crystal structure search calculations based on first-principles method. The structural stability, mechanical, electronic and superconducting properties of the stable phases have been discussed. The predicted five compounds BeB$_2$Ta, BeB$_3$Ta$_2$, BeBTa, BeBTa$_2$, and Be$_2$B$_2$Ta have been found to be highly dense and very hard materials. They may be potential candidates as superhard materials in different technological applications. All these ternary stable compounds are metallic and spin-orbit coupling (SOC) effect splits some energy bands arisen from Ta 5d. In $BeB_3Ta_2$, $BeBTa$, and $BeBTa_2$, Ta 5d and B 2p orbitals are strongly hybridized while Be 2p orbitals have significant contribution to the hybridization in $BeB_2Ta$ and $Be_2B_2Ta$. Thus, only two of them ($BeB_2Ta$ and $Be_2B_2Ta$) have been found to be superconductors within Migdal-Eliashberg theory. Therefore, any stable combination of B-rich with electron-rich transition metal ( for example, La, Hf, etc.) may be potential superconductor if the hybridization between B-2p and electron-rich element-d orbitals is disturbed by another 2s-element (such as Mg, Ca, Sr, etc.). The calculated critical temperature including SOC effect is 8 and 9 K for $BeB_2Ta$ and $Be_2B_2Ta$, respectively. Because of their energetic and dynamic stability, these compounds might be favorable to synthesize in the laboratory.

## 4. Computational Methodology

The thermodynamically stable crystal structure searches for ternary Be-B-Ta were carried out with first-principles evolutionary algorithm implemented in USPEX code (Universal Structure Predictor: Evolutionary Xtallography) [59–62]. This method of crystal structure search has been successfully applied in many systems (e.g., Refs. [63–66]). The calculation of lowest enthalpy phase of a given element composition was performed with the combination of USPEX and Quantum Espresso (for structural relaxation) [67]. At ambient condition, two separate variable composition ternary searches were carried out for 3-6 atoms per unit-cell and 6-12 atoms/unit-cell with the hope that no potential structures were missed out. When variable composition searches were finished, the structure searches at fixed composition predicted to be stable in the previous calculations were carried out with up to 22 atoms/unit-cell to find the lowest enthalpy space group structure for each composition. The relaxation performed during the structure search by using Perdew-Burke-Ernzerhof (PBE) generalized gradient approximation (GGA) [68,69]. For this, Vanderbilt plane-wave pseudopotentials [70], 350 eV cutoff energy for wavefunction, and 0.8-0.02 Å$^{-1}$ k-point for BZ integration were used. During the variable composition calculation, USPEX was set to produce 20 generations. Each generation contained 150 individual structures and all structures in the first generation was produced randomly. In all other generations, 20% individuals were produced through random based on space group symmetry, 40% individuals of one generation were applied to produce individuals of next generations by heredity, 20% generated through soft mutations, and 20% individuals were produced through transmutation. After the stable structure obtained, the structural relaxation was performed with 520 eV cutoff energy and 0.003 Å$^{-1}$ k-point resolution to calculate formation enthalpy and lattice parameters. The vc-relax convergence criteria were set to $1 \times 10^{-8}$ eV/atom for energy, and 0.0001 eV/Å for force on each atom. The best convergence was obtained by using Thomas-Fermi (TF) charge mixing mode [71].

By using strain-stress approach, elastic constants were calculated with CASTEP code [72]. The same pseudopotentials, functional, cutoff energy, and k-point were used in this calculation. The bulk modulus (B) and shear modulus (G) were obtained by using the calculated elastic constants within Voigt-Reuss-Hill scheme [73]. Chen's formula was used to calculate Vickers hardness [74].

Electronic structure, phonon dispersion, the density of states, and superconducting transition temperature $T_c$ were calculated using Quantum Espresso. In the electron-phonon coupling (EPC) parameter calculation, first Brillouin zone was sampled using a $2 \times 4 \times 4$, $4 \times 4 \times 4$, $4 \times 4 \times 4$, $4 \times 4 \times 2$, and $4 \times 4 \times 4$ q-point mesh and $4 \times 8 \times 8$, $8 \times 8 \times 8$, $8 \times 8 \times 8$, $8 \times 8 \times 4$, and $8 \times 8 \times 8$ k-point mesh for BeB$_2$Ta, BeB$_3$Ta$_2$, BeBTa, BeBTa$_2$, and Be$_2$B$_2$Ta, respectively. The Fermi surface evaluation was performed using a two times denser k-point mesh of the above. Besides these parameters, Marzari-Vanderbilt smearing [75] of width 0.03 Ry was used in the BZ sampling. The modified Allen-Dynes equation was used to evaluate Tc with the most widely accepted value of μ* (0.1) [56]. For spin-orbit calculation, fully-relativistic pesudopotential of Ta was used [76].

# Supporting Information

Prediction of fundamental properties of Be-B-Ta based novel ternary compounds from first-principles calculations

Enamul Haque[1*], M. Anwar Hossain[1], and Catherine Stampfl[2]

[1] Department of Physics, Mawlana Bhashani Science and Technology University, Santosh, Tangail-1902, Bangladesh

[2] School of Physics, The University of Sydney, Sydney, New South Wales, 2006, Australia.

*enamul.phy15@yahoo.com

This supporting document contains structural information and Fermi surfaces of all stable ternary compounds.

**Structural Information:**

Table-SI 1: Fully relaxed fractional atomic coordinates and the corresponding Wyckoff symbol of all the studied stable-compounds.

| Compound | Space group | Wyckoff symbol | Atom | x | y | z |
|---|---|---|---|---|---|---|
| $BeB_2Ta$ | P2/m(c) (#10) | 1e | Be | 0.00000 | 0.50000 | 0.50000 |
| | | 1a | Be | 0.00000 | 0.00000 | 0.00000 |
| | | 2m | B | 0.80993 | 0.40445 | 0.00000 |
| | | | B | 0.19007 | 0.59555 | 0.00000 |
| | | 2m | B | 0.43907 | 0.71867 | 0.00000 |
| | | | B | 0.56093 | 0.28133 | 0.00000 |
| | | 2n | Ta | 0.30565 | 0.15232 | 0.50000 |
| | | | Ta | 0.69435 | 0.84768 | 0.50000 |
| $BeB_3Ta_2$ | Imm2 (#44) | 2a | Be | 0.0000 | 0.00000 | 0.0000 |
| | | 2a | B | 0.0000 | 0.00000 | 0.33946 |
| | | 2b | B | 0.50000 | 0.00000 | 0.32234 |
| | | 2b | B | 0.00000 | 0.50000 | 0.49885 |
| | | 4c | Ta | 0.25092 | 0.50000 | 0.16576 |
| $BeBTa_2$ | Amm2(#38) | 2a | Be | 0.00000 | 0.00000 | 0.0000 |
| | | 2b | B | 0.50000 | 0.00000 | 0.12600 |
| | | 2b | Ta | 0.50000 | 0.00000 | 0.41322 |
| | | 2a | Ta | 0.00000 | 0.50000 | 0.20664 |
| $BeBTa$ | $P\bar{6}m2$ (#187) | 1a | Be | 0.00000 | 0.00000 | 0.00000 |
| | | 1c | B | 0.33333 | 0.66667 | 0.00000 |
| | | 1f | Ta | 0.66667 | 0.33333 | 0.50000 |
| $Be_2B_2Ta$ | I4/mmm (#139) | 4d | Be | 0.50000 | 0.00000 | 0.25000 |
| | | 4e | B | 0.00000 | 0.00000 | 0.60217 |
| | | 2a | Ta | 0.00000 | 0.00000 | 0.00000 |

The fully relaxed fractional atomic coordinates of predicted five ternary compounds are listed in Table-SI 1. The corresponding Wyckoff symbol of these atomic coordinates is also presented. The symmetrized crystallographic information files (CIF) of these compounds are also included in this submission.

**Fermi surface:**

The calculated electronic structure and superconducting transition temperature predict that only B or both B and Be rich compounds (BeB$_2$Ta and Be2B$_2$Ta) shows superconductivity. In Ta-rich compounds, Ta 5d and B 2p orbitals are strongly hybridized that favors the absence of superconductivity. For further support, the calculated distribution of Fermi velocity of electronic states on the Fermi surface is shown in Fig. SI-1 with a color plot. High, middle and low values of Fermi velocity are indicated by the red, green, and blue colored region indicates, respectively.

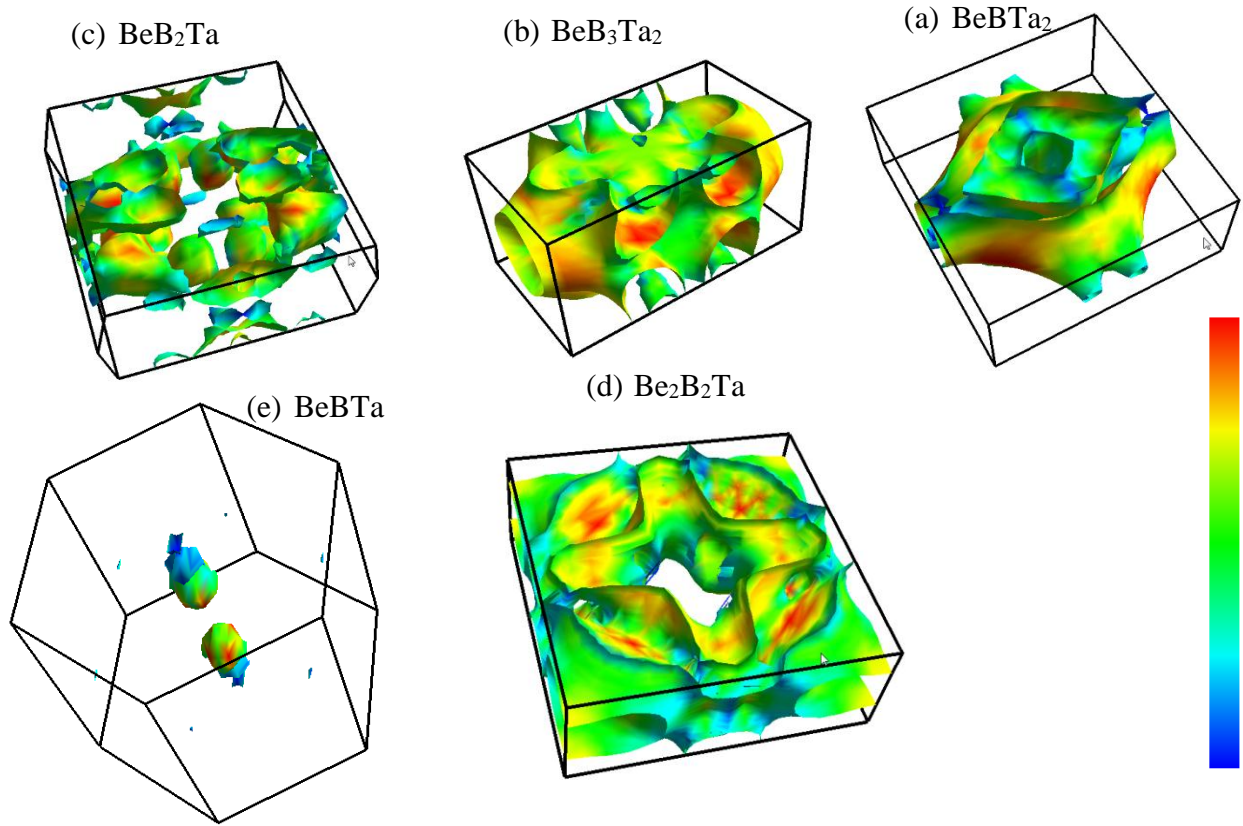

Fig. SI-1: Calculated Fermi velocity of the electronic states on Fermi surface of: (a) BeB$_2$Ta, (b) BeB$_3$Ta$_2$, (c) BeBTa$_2$, (d) BeBTa, and (e) Be$_2$B$_2$Ta. High, middle and low values of Fermi velocity are indicated by the red, green, and blue colored region indicates, respectively.

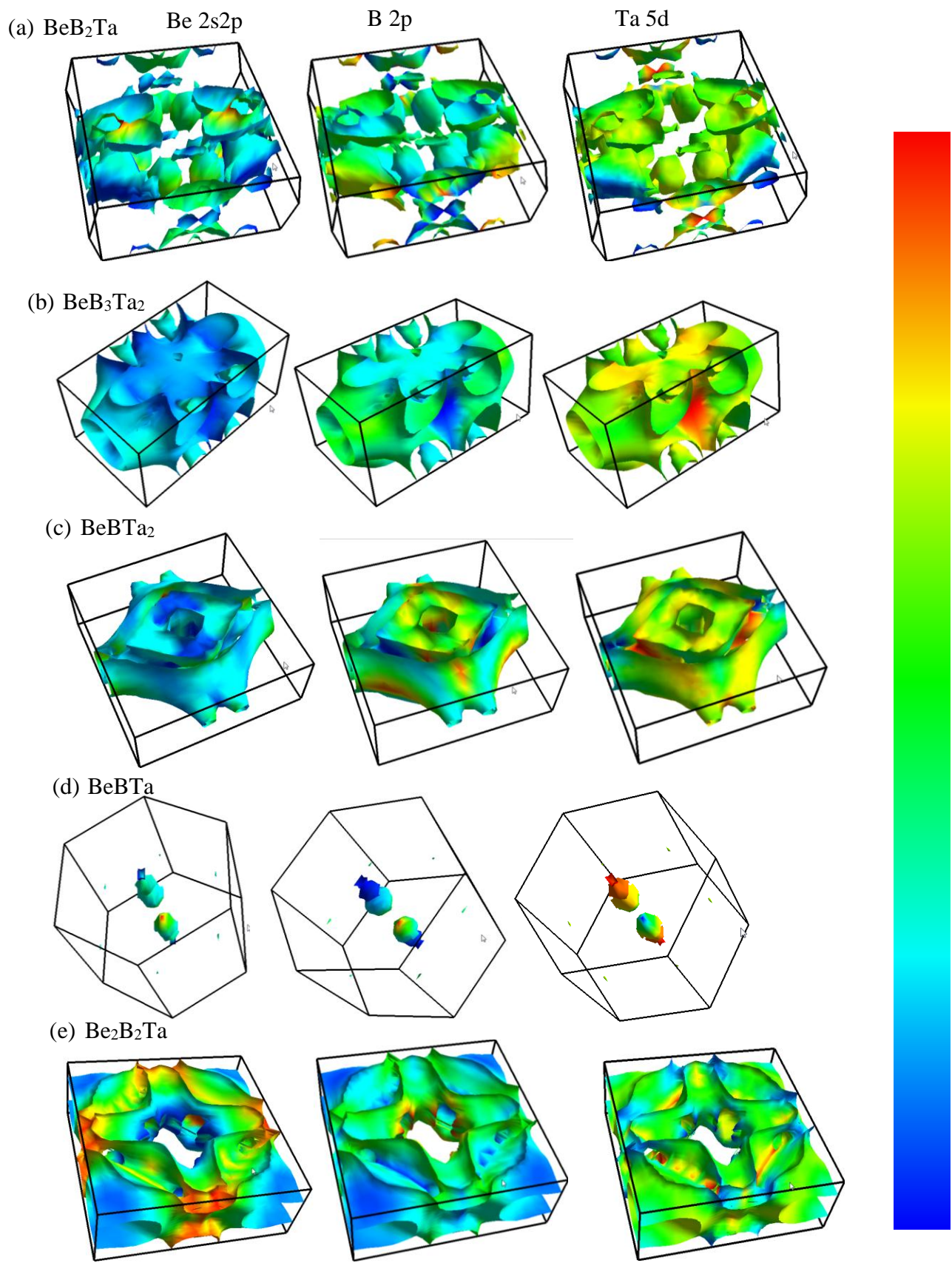

Fig. SI-2: Calculated projection of atomic orbitals Be 2s2p, B 2p, and Ta 5d on Fermi surface of: (a) $BeB_2Ta$, (b) $BeB_3Ta_2$, (c) $BeBTa_2$, (d) BeBTa, and (e) $Be_2B_2Ta$.

The Fermi velocity of electronic states on the Fermi surface of the Ta-rich compound is much lower than that of either B or both Be and B rich compounds (as indicated by red colored region). The Fermi surfaces were drawn by using the FermiSurfer [1]. BeBTa has very small Fermi surface, like a carrier deficient compound. To get further information, the calculated projections of the atomic orbitals Be 2s2p, B 2p and Ta 5d to the electronic states on the Fermi surface of these compounds are shown in Fig. SI-2. Be 2s2p and B 2p orbitals have no dominating regions in Ta-rich compounds while they have significant contribution in $BeB_2Ta$ and $Be_2B_2Ta$. By comparing Fig. SI-1 and Fig. SI-2, the Fermi velocity is low in the regions where Ta 5d states are dominant. The small Fermi surface of BeBTa indicates the deficiency of carriers and hence, the absence of superconductivity. These results give further support that only B-rich/Be-B-rich compounds may show superconductivity.

**References**

[1] http://fermisurfer.osdn.jp/.